\documentclass[aps,prl,10pt,superscriptaddress,twocolumn]{revtex4-1}

\usepackage{amsmath,amssymb,mathrsfs}
\usepackage{graphicx}
\usepackage{bm}

\DeclareMathOperator{\tr}{tr}

\def\Tr{\mbox{Tr}\,}
\newcommand{\la}{\label}
\newcommand{\bbm}{\begin{multline}}
\newcommand{\eem}{\end{multline}}
\newcommand{\be}{\begin{equation}}
\newcommand{\ee}{\end{equation}}

\newcommand{\bea}{\begin{eqnarray}}
\newcommand{\eea}{\end{eqnarray}}
\newcommand{\p}{\partial}

\newcommand{\sgn} {\mbox{sgn}\,}

\newcommand{\comment}[1]{}
\usepackage{color}

\allowdisplaybreaks[1]

\begin{document}

\title{Framing Anomaly in the Effective Theory of Fractional Quantum Hall Effect}

\author{Andrey~Gromov}
\affiliation{Department of Physics and Astronomy, Stony Brook University,  Stony Brook, NY 11794, USA}

\author{Gil~Young~Cho}
\affiliation{Department of Physics and Institute for Condensed Matter Theory, University of Illinois, 1110 W. Green St., Urbana, Illinois 61801-3080, USA}
\author{Yizhi~You}
\affiliation{Department of Physics and Institute for Condensed Matter Theory, University of Illinois, 1110 W. Green St., Urbana, Illinois 61801-3080, USA}

\author{Alexander G.~Abanov}
\affiliation{Department of Physics and Astronomy, Stony Brook University,  Stony Brook, NY 11794, USA}
\affiliation{Simons Center for Geometry and Physics,
Stony Brook University,  Stony Brook, NY 11794, USA}

\author{Eduardo~Fradkin}
\affiliation{Department of Physics and Institute for Condensed Matter Theory, University of Illinois, 1110 W. Green St., Urbana, Illinois 61801-3080, USA}
\affiliation{Kavli Institute for Theoretical Physics, University of California Santa Barbara, CA 93106-4030, USA}

\date{\today}

\begin{abstract}
We consider the geometric part of the effective action for Fractional Quantum Hall Effect (FQHE). It is shown that accounting for the framing anomaly of the quantum Chern-Simons theory is essential to obtain the correct gravitational linear response functions. In the lowest order in gradients the linear response generating functional includes Chern-Simons, Wen-Zee and gravitational Chern-Simons terms. The latter term has a contribution from the framing anomaly which fixes the value of thermal Hall conductivity and contributes to the Hall viscosity of the FQH states on a sphere. We also discuss the effects of the framing anomaly on linear responses for non-Abelian FQH states.
 \end{abstract}


\maketitle


Fractional quantum Hall (FQH) states exemplify genuinely new states of matter with long range topological order. These states owe their fascinating properties to the strong interaction between the electrons partially filling one or several Landau levels. Although much is known about general properties of the FQH states, the general problem of strongly interacting electrons in quantizing magnetic field defies controlled analytical treatment. To date FQH states are the prototype of a topological quantum fluid.

In addition to a quantized topological electromagnetic response, FQH states (as well as general $2+1D$ topological phases with broken time-reversal symmetry) exhibit other geometric responses such as Hall viscosity \cite{1995-AvronSeilerZograf,levay1995berry,2009-Read-HallViscosity,2011-ReadRezayi} and thermal Hall conductance \cite{Kane-Fisher-LR,2000-ReadGreen,CappelliLR}. These responses can be computed via adiabatic arguments on a torus,  directly from the Laughlin wave function on a curved space \cite{douglas2010bergman, klevtsov2014random, CLW}, using Chern-Simons gauge theory or the projective parton construction \cite{2014-ChoYouFradkin}. Placing the FQH states on a curved manifold proved to be a useful tool as it allowed to probe more response functions \cite{1993-frohlich,WenZeeShiftPaper} and, therefore, distinguish FQH states having identical charge response \cite{WenZeeShiftPaper}.

Following the elegant construction of Ref. \cite{WenZeeShiftPaper,wen-Kmatrix} we introduce the effective action of a general Abelian FQH state on a curved state as
\be
	S = -\frac{1}{4\pi} \int  \;\Big[K_{IJ} a^I d a^J + 2q_I Ada^I + 2s_I\omega da^I \Big]\;.
 \la{WZaction}
\ee
Here we use concise ``form notation'' so that $Ada^I\leftrightarrow\epsilon^{\mu\nu\lambda} A_\mu\p_\nu a^I_\lambda \,d^3x$, etc., and integration in Eq.\eqref{WZaction} is taken over three-dimensional space-time. The theory contains $\kappa$ hydrodynamic gauge fields $a^I$, $I=1,\ldots,M$ coupled to external electro-magnetic vector potential $A_\mu$ and to external geometry through the abelian $SO(2)$ spin connection $\omega_\mu$.  $K_{IJ}$ is the (symmetric) $M \times M$ $K$-matrix, $q_I$ is the charge vector and $s_J$ is the spin vector \cite{WenZeeShiftPaper}. We will also use bold symbols for matrices and vectors, so that ${\bf K}$, ${\bf q}$ and ${\bf s}$ denote a $K$-matrix, charge vector and spin vector correspondingly. 

The action of Eq.\eqref{WZaction} describes interactions of the conserved currents $j^I_\mu\equiv \frac{1}{2\pi}\epsilon^{\mu\nu\lambda}\p_\mu a_\lambda^I$ with the external gauge field $A_\mu$ and the  background geometry of the spatial manifold parametrized by the Abelian spin connection $\omega_\mu$.

Two remarks are in order. First, only the leading terms in the gradient expansion are kept  in Eq.\eqref{WZaction}. The first term in Eq.\eqref{WZaction} is the action of a Chern-Simons gauge theory with  gauge group $U(1)^M$. This term is independent of the metric and, up to some caveats discussed below,  it is the topological
part of the action. The higher gradient terms are, of course, present for any FQH state but are suppressed by the gap in the spectrum. Second, as these leading orders
are written in terms of differential forms it is clear that the action Eq.\eqref{WZaction} does not depend on the metric of the background other than through the spin connection $\omega$ in the last term of Eq.\eqref{WZaction}. 

The simplest quantum Hall state is described by a $1 \times 1$ $K$-matrix, ${\bf K}=1$, with ${\bf q}=1$ and ${\bf s}=1/2$ as charge and spin ``vectors'' respectively. It corresponds to the filling factor $\nu=1$, i.e., the spineless (spin polarized) electrons  filling up the lowest Landau level. In this case, the action Eq.\eqref{WZaction} reduces  to
\be
	S = -\frac{1}{4\pi} \int  \Big[ a d a + 2 Ada + \omega da\Big] \,.
 \la{WZaction1}
\ee
To find the electromagnetic and gravitational linear response functions from Eq.\eqref{WZaction1} one has to integrate out the hydrodynamic gauge field $a_\mu$ to find the generating functional
for linear responses. A traditional way of treating this integration \cite{WenZeeShiftPaper} is to substitute the solution of saddle point equation for $a_\mu$ following from Eq.\eqref{WZaction1}, $a_\mu=-A_\mu-\omega_\mu/2$, back into Eq.\eqref{WZaction1}, and obtain
\be
	S_{\rm eff}' = \frac{1}{4\pi} \int  \left(A+\frac{1}{2}\omega\right) d \left(A+\frac{1}{2}\omega\right) \,.
 \la{seffprime1}
\ee
However, Eq.\eqref{seffprime1} does not agree with the result of direct computation \cite{Abanov-2014} of the effective action for non-interacting fermions at $\nu=1$. The result found in Ref. \cite{Abanov-2014} is
\be
	S_{\rm eff} = \frac{1}{4\pi} \int \left(A+\frac{1}{2}\omega\right) d \left(A+\frac{1}{2}\omega\right) 
	-\frac{1}{48\pi}\int \omega d\omega \,.
 \la{seffcorr1}
\ee
Eq.\eqref{seffcorr1} and Eq.\eqref{seffprime1} differ by the additional gravitational Chern-Simons term. 

In a recent publication \cite{2014-ChoYouFradkin}, three of us used Chern-Simons gauge theory (to represent flux attachment) and projective parton constructions to derive, from microscopic models, the effective actions of the hydrodynamic fields and their coupling to the background geometry. A key ingredient of this construction is that the worldlines of these composite particles are always framed and, as a result, the effective action yields the correct values of the couplings of the Wen-Zee term and of the Hall viscosity. However, a consistent theory of the gravitational Chern-Simons term was lacking.
Below we explain that the appearance of this term is not accidental, but is a consequence of a general phenomenon that is present in the quantum Chern-Simons theory known as the framing anomaly \cite{witten1989quantum,WittenBarNatan}. We will see that the framing anomaly is the key ingredient to obtain a consistent effective theory for all FQH states.

\paragraph{Main results.}

In this work we generalize the action of Eq.\eqref{seffcorr1} to arbitrary Abelian and non-Abelian FQH states, providing a generating functional in the leading order in derivatives.
For general Abelian FQH states coupled to external electromagnetic field and geometry defined by Eq.\eqref{WZaction} the 
topological part of the effective action is given by
\bea
	S_{\rm eff} &=& S_{K} +S_{\rm anom} \,,
 \la{seffcorr} \\
	S_{K} &=&  \frac{1}{4\pi} \int \left(\bm{q}^T A +\bm{s}^T \omega \right) 
	\bm{K}^{-1} d \left(\bm{q}A+\bm{s}\omega\right)\,,
 \la{sK} \\
 	S_{\rm anom} &=& - \frac{c}{96\pi} \int \tr\left(\Gamma d\Gamma + \frac{2}{3}\Gamma^3\right)\,,
 \la{sanom}
\eea
where we used matrix notations for $K$-matrix, spin and charge vectors. 
The contribution shown in Eq.\eqref{sanom} is the framing anomaly of quantum Chern-Simons theory 
\cite{witten1989quantum,WittenBarNatan}. The coefficient $c$ is the chiral central charge which, for a general Abelian theory,  is equal to 
\be
	c = 
	\sgn \bm{K} \,,
\ee
where $\sgn {\bm K}=N_+ - N_-$ is the signature of the $K$-matrix.  $N_\pm$ is the number of positive (negative) eigenvalues of the $K$ matrix.
Then, as it will be explained below, the general formula of Eq.\eqref{sanom} reduces to 
\be
	S_{\rm anom} = - \frac{c}{48\pi} \int \omega d\omega \,,
 \la{sanomomega}
\ee
for a particular choice of geometric background. As a check, it is easy to see that Eq.\eqref{sanomomega} reduces to the last term of Eq.\eqref{seffcorr1} for ${\bf K}=1$.

Using the projective parton approach \cite{Barkeshli2010}, we can also find the generalization of Eq.\eqref{seffcorr}-Eq.\eqref{sanom} for non-Abelian FQH states such as ${\mathbb Z}_k$ parafermion states. The only yet crucial difference here from the Abelian states is that the central charges of the non-Abelian states are {\em rational fractions}, instead of  an integer. More precisely, the chiral edge theories of the non-Abelian states are the $G/H$-coset conformal field theory (CFT) whose central charge is $c_{G/H} = c_{G} - c_{H}$ where     
\be\la{NA-charge}
c_{G} = \frac{k \dim(G)}{k+h}
\ee
is  a rational number. In Ref.~\cite{2014-ChoYouFradkin} it was noted that a naive calculation of the gravitational anomaly term of  Eq.\eqref{sanomomega} using the projective parton construction yields an incorrect (integer) value for the chiral central charge.
In this Letter, we  show that the framing anomaly, which was missing in the work \cite{2014-ChoYouFradkin}, yields in all cases the correct value of the chiral central charge. 

\paragraph{Geometric responses.}

Let us relate the contribution of the framing anomaly to the effective action to physical observables.
We focus here on various geometric response functions. These response functions are known to be of interest in the physics of FQHE and have been studied previously. They include the thermal Hall conductance  \cite{Kane-Fisher-LR,2000-ReadGreen,CappelliLR} and the Hall viscosity \cite{1995-AvronSeilerZograf,levay1995berry,2009-Read-HallViscosity,2011-ReadRezayi}.

The framing anomaly contribution to the effective action can be considered as the bulk manifestation of the thermal Hall conductance $\kappa_H$ 
which, for a quantum Hall fluid,  is known to be proportional to the chiral central charge of the chiral edge states of the FQH fluid \cite{Kane-Fisher-LR,2000-ReadGreen,CappelliLR}
\be
	\kappa_H = c \frac{\pi k_B^2 T}{6}\, .
 \la{kappaH}
\ee
where $c$ is the central charge.
On the other hand, in the presence of the background curvature the gravitational Chern-Simons term also contributes to the Hall viscosity. If the quantum Hall state is  on a sphere of constant Ricci curvature $R$, then the Hall viscosity is given by \cite{Abanov-2014,gromov2014density}
\be
 \la{Hall-visc}
	\eta_H = \frac{\bar s}{2} n - \frac{c}{24}\frac{R}{4\pi}\, . 
\ee
The last term in Eq.\eqref{Hall-visc} is a finite size correction to the well known relation $\eta_H = \frac{\bar s}{2} n$. The appearance of the chiral central charge $c$ is, therefore, very natural. We should note that the gravitational Chern-Simons term does {\it not} describe the {\it bulk} thermal Hall effect \cite{Stone-Gravitational}. The latter can be understood as a response to the geometry with temporal torsion and is not topologically protected \cite{bradlyn2014low,GA-thermal}.

\paragraph{Framing anomaly.}

Before proceeding to applications of general results to particular FQH states we give a very brief review of the framing anomaly tailored to our purposes.
The integration over the hydrodynamic Chern-Simons gauge field in the action of the type Eq.\eqref{WZaction1} is done by substituting the solutions of equations of motion back into the action. While it is true that stationary phase approximation for the gaussian integral is exact there is a subtlety that arises when Chern-Simons theory is defined on a curved space.
  
It is well known that the Chern-Simons theory is {\it topological } at the classical level, {\it i.e.} it does not depend on the metric and has vanishing stress-energy tensor. However, this is not  true for the full quantum theory \cite{witten1989quantum,WittenBarNatan}. The reason is that while the action is metric-independent, the path integral measure does depend on metric in a non-trivial way. Indeed, the definition of the path integral measure $\mathcal{D}\mathcal A$ requires gauge fixing, which should be defined in a covariant way to avoid dependence of the partition function on the choice of  coordinates. 
For example, the gauge fixing can be done by including an additional gauge fixing term into the action
\be
 \la{gaugefix}
	S_{\phi} = \int dV \phi D^\mu \mathcal A_\mu\, ,
\ee 
with the integration over the auxiliary field $\phi$ included in the path integral. The term Eq.\eqref{gaugefix} depends on the {\it geometry} of the manifold through both covariant derivative $D^\mu$ and the invariant space-time integration measure $dV$. The term of Eq.\eqref{gaugefix} is understood as a part of the definition of the integration measure  $\mathcal{D}\mathcal{A}_\mu$. 

The dependence of the full partition function $Z$ on the metric of the manifold can be quantified \cite{witten1989quantum,WittenBarNatan}. Consider the partition function of the Chern-Simons theory with arbitrary compact, semi-simple group $G$ at level $k$. Its partition function is given by\cite{witten1989quantum}
\bea
	Z&=& \int \mathcal D\mathcal A \mathcal{D}\phi \, \exp\left\{-i\frac{k}{4\pi}\int_M \tr \left( \mathcal A d\mathcal A 
	+ \frac{2}{3} \mathcal A^3 \right)- i S_{\phi} \right\} 
 \nonumber \\
	&=& \tau \exp \left\{-i \frac{c}{96\pi}\int_M \tr \left(\Omega d\Omega 
	+ \frac{2}{3}\Omega^3 \right)\right\} \,,
 \la{witten1}
\eea
where $\tau$ is the Ray-Singer analytic torsion \cite{schwarz1978partition}. The latter is a topological invariant and is not important for the upcoming discussion. The phase of the partition function $Z$ is given by the framing anomaly and $c$ is the chiral central charge given by Eq.\eqref{NA-charge}.

In Eq.\eqref{witten1} $\Omega^{a}{}_{b,\mu}$ is the Levi-Civita $SO(1,2)$ spin connection \cite{nakahara2003geometry}. We denote it by $\Omega$ to avoid the confusion with the $SO(2)$ spin connection $\omega$ (see below). In this work we are interested in quantum Hall states, which are inherently non-relativistic systems. For this reason we turn off the temporal components of the spin connection $\Omega^{a}{}_{0,\mu}=\Omega^{0}{}_{b,\mu}=0$ because non-relativistic physical systems generally do not couple to these components. With this choice the $SO(2)$ component of the spin connection $\omega_\mu\equiv \Omega^1{}_{2,\mu}$ is precisely the one used in Eq.\eqref{WZaction}. Then, we obtain
\be
 \la{NC-gCS}
	\frac{c}{96\pi}\int_M \tr \left(\Omega d\Omega + \frac{2}{3}\Omega^3\right) 
	= \frac{c}{48\pi}\int_M \omega d\omega \,.
\ee

\paragraph{Relation to the gravitational anomaly.}

Here we emphasize the relation of the framing anomaly to the edge theory of FQHE. The edge theory has a contribution from the gravitational anomaly \cite{CardyLR,affleckLR} which can be related to the bulk gravitational Chern-Simons term in the following way.
First, let us rewrite the gravitational Chern-Simons term Eq.\eqref{NC-gCS} replacing the $SO(1,2)$ spin connection $\Omega$ by Christoffel symbols as \cite{chamseddine1992two}
\bea
	\frac{c}{96\pi} \int \tr \left(\Omega d\Omega + \frac{2}{3}\Omega^3\right) 
	&=&  \frac{c}{96\pi} \int \tr \left(\Gamma d\Gamma + \frac{2}{3}\Gamma^3\right) 
 \nonumber \\
	&-& \frac{c}{288\pi}\int \tr(e^{-1}de)^3 \,,
 \la{G-O}
\eea
The last term in this relation describes the winding number of the dreibeins $e$ and is irrelevant here since the variations of this term on a closed manifold vanish \cite{Note}. 

The gravitational Chern-Simons term written in terms of Christoffel symbols $\Gamma^\mu{}_{\nu,\rho}$ is not invariant with respect to changes of coordinates in the presence of a boundary and induces the gravitational anomaly of the edge theory \cite{callan1985anomalies}. 
Thus, in general expressions such as Eq.\eqref{sanom}, we present the contributions of the framing anomaly in terms of Christoffel symbols to emphasize the relation to the gravitational anomaly and, in turn, to the thermal Hall effect \cite{2000-ReadGreen}.

\paragraph{Effective action for  Abelian FQH  states.}

The effective action for a general Abelian FQH state can be written as 
\bea
	S_{eff} &=& \frac{\nu}{4\pi} \int   
	\Big((A + \bar s\omega)d(A + \bar s\omega)+\beta \omega d\omega \Big)   
 \nonumber \\
	&-& \frac{c}{96\pi} \int \tr \left[\Gamma d \Gamma + \frac{2}{3} \Gamma^3 \right]\, ,
 \la{seffgen}
\eea
where $\nu$ is the filling fraction, 
$\bar{s}$ is the average orbital spin, $\beta=\nu^s-\nu\bar{s}^{2}$ is the \emph{orbital spin variance}, and $\nu^s$ is the ``spin filling fraction'' \cite{WenZeeShiftPaper}, given by
\begin{equation}
	\nu = \bm{q}^T \bm{K}^{-1} \bm{q},\quad \nu \bar{s}
	= \bm{q}^T \bm{K}^{-1} \bm{s},\quad \nu^s = \bm{s}^T \bm{K}^{-1} \bm{s}\, .
\end{equation}
For the Laughlin series at the filling $\nu = \frac{1}{2r+1}$ we have 
\be
 \la{Laughlin}
	\bar{s} = r+\frac{1}{2}\, , \qquad \beta = 0\, , \qquad c =1 \,.
\ee
The $K$-matrix for the Jain series can be found in Ref. \cite{wen-Kmatrix}. For the Jain series at the filling $\nu = \frac{p}{2rp\pm 1}$ (with $p,r\in \mathbb{Z}$ and $p\geq 1$, $r\geq 1$) we have 
\be
 \la{Jain1}
	\bar{s} = \pm r +\frac{p}{2}\, , \quad \beta =\pm\frac{p(p^2-1)}{12}\, ,
	\quad c = 1\pm(p-1) \,.
\ee
The relations Eqs.\eqref{Laughlin}-\eqref{Jain1} can be derived through the flux attachment procedure \cite{LopezFradkin,2014-ChoYouFradkin} or  by the projective parton construction \cite{Wen1999}. One can use Eqs.\eqref{Laughlin}-Eq.\eqref{Jain1} to compute Hall viscosity and thermal Hall conductivity from Eqs.\eqref{kappaH}-\eqref{Hall-visc}.

\paragraph{Non-Abelian states.}

In the following we will derive the effective action for the non-abelian ${\mathbb Z}_{k}$ Read-Rezayi (RR) parafermion states \cite{Read1999} at filling $\nu = \frac{k}{Mk+2}$. While the problem of deriving the bulk effective theory for a generic non-abelian gapped FQH state is not solved, the answer for a variety of different states can be obtained through the parton construction \cite{Wen1999,Barkeshli2010}. The effective bulk theory for the non-abelian ${\mathbb Z}_{k}$ Read-Rezayi parafermion states at filling $\nu = \frac{k}{Mk+2}$ is given by the $(U(M)\times Sp(2k))_1$ Chern-Simons theory \cite{Barkeshli2010} and $U(1)_1^{2k+M}$ Abelian theory.
\bea
	S &=&\frac{1}{4\pi} \int \tr\left[ada + \frac{2}{3} a^3 +  \omega d a\right] 
 \nonumber \\ 
	&-&  \frac{1}{4\pi} \int \tr \Big[ bdb+2(QA+S\omega)db\Big]\,,
 \la{NA}
\eea
where $Q=\frac{1}{kM+2}\mbox{diag}\,(1_{2k},k\times 1_{M})$ and $S=\frac{1}{2}1_{2k+M}$ are $(2k+M)\times (2k+M)$ charge and spin matrices. There are $2k+M$ hydrodynamic $U(1)$ gauge fields $b$ and one non-abelian $U(M)\times Sp(2k)$ field $a$. In the second line of Eq. \eqref{NA} we have coupled the bulk theory to external electromagnetic field and geometry as in Eq. \eqref{WZaction} (see \cite{2014-ChoYouFradkin}). In  Eq.\eqref{NA} we have essentially used the coset construction of \cite{moore1989taming}. Note that the introduction of the abelian fields $b$ does not change the degeneracy on the higher genus surfaces because the corresponding $K$-matrix is unity.

Integration over the low energy degrees of freedom implies the universal effective action Eq.\eqref{seffgen} with the filling factor, the average orbital spin, and the orbital spin variance given by
\bea
	\nu &=& \Tr Q^{2}= \frac{k}{Mk+2}\,, 
 \\
	\bar{s} &=& \nu^{-1}\Tr QS=\frac{M+2}{2}\,, 
 \\
	\beta &=& 0\,.
\eea
The chiral central charge  $c$ of the boundary $U(1)_1^{2k+M}/(U(M)\times Sp(2k))_1$ coset CFT is given by 
\be
	c=c_{U(1)_{1}^{2k+M}}-c_{U(M)_{1}}-c_{Sp(2k)_{1}}=\frac{3k}{k+2} \,,
\ee
which is the correct value of the central charge of the edge states of the RR parafermion states.

{\it Note:} In this version of the paper we have added the correct versions of Eq. (21) and Eq. (24) which are incorrect in the original posting of the paper and in the published version. The reasons for the change are given explicitly in the Erratum provided in the end of the manuscript.

\paragraph{Conclusions.}

We have derived the effective action for arbitrary FQH states on a curved manifold. It turned out to be very important that quantum Chern-Simons theory depends on the metric through the measure of the functional integral. This metric dependence ultimately leads to an additional gravitational Chern-Simons term in the effective action that fixes the value of thermal Hall conductivity and the ``finite size'' correction to the Hall viscosity. We have derived the effective action for the various abelian states and also found complete agreement with previously known results.

A.G. is grateful to the hospitality and inspirational atmosphere of the Les Houches Summer School on Topological Aspects of Condensed Matter Physics. EF thanks the KITP (and the Simons Foundation) and the IRONIC14 program for support and hospitality. EF thanks C. Nayak, T. Hughes, S. Ryu, and X.-G. Wen  for  discussions.  This work of  was supported in part by the NSF grants No. DMR-1206790 at Stony Brook University (A.G.A.), DMR-1064319 (GYC,EF), DMR 1408713 (YY,EF) at the University of Illinois, and PHY11-25915 at KITP (EF).

\bibliographystyle{my-refs}




\bibliography{Bibliography}

\begin{thebibliography}{10}
\bibitem{Note}
Importantly, the replacement of the $SO(2)$ spin connection of the $S_K$ part
  of the action Eq.\eqref{sK} by Christoffel symbols cannot be done.
\end{thebibliography}

\end{document}


\title{Erratum: Framing Anomaly in the Effective Theory of Fractional Quantum Hall Effect, 
[Phys. Rev. Lett., \textbf{114}, 016805 (2015)]}

\author{Andrey~Gromov}
\affiliation{Department of Physics and Astronomy, Stony Brook University,  Stony Brook, NY 11794, USA}
\author{Gil~Young~Cho}
\affiliation{Department of Physics and Institute for Condensed Matter Theory, University of Illinois, 1110 W. Green St., Urbana, Illinois 61801-3080, USA}
\affiliation{Department of Physics, Korea Advanced Institute of Science and Technology, Daejeon 305-701, Korea}
\author{Yizhi~You}
\affiliation{Department of Physics and Institute for Condensed Matter Theory, University of Illinois, 1110 W. Green St., Urbana, Illinois 61801-3080, USA}
\author{Alexander G.~Abanov}
\affiliation{Department of Physics and Astronomy, Stony Brook University,  Stony Brook, NY 11794, USA}
\affiliation{Simons Center for Geometry and Physics,
Stony Brook University,  Stony Brook, NY 11794, USA}
\author{Eduardo~Fradkin}
\affiliation{Department of Physics and Institute for Condensed Matter Theory, University of Illinois, 1110 W. Green St., Urbana, Illinois 61801-3080, USA}
\affiliation{Kavli Institute for Theoretical Physics, University of California Santa Barbara, CA 93106-4030, USA}

\date{\today}
\maketitle

In our recent published work \cite{Gromov2015}  there is an error in the computation of the geometric response functional for the ${\mathbb Z}_k$ parafermion fractional quantum Hall (FQH) states \cite{Read1999} at the filling $\nu = \frac{k}{Mk+2}$ which  lead to the incorrect result that  the spin variance $\beta$ for the states with $M > 0$ and $k >1$ does not vanish. The error  originated from the incorrect form of the effective theory given by Eq.(21) of  Ref. \cite{Gromov2015}, which misses an additional allowed coupling to spin connection in the action of Eq. (21) whose correct form is
\begin{equation*}
	S =\frac{1}{4\pi} \int \tr\left[ada + \frac{2}{3} a^3 +  \omega d a\right] - \frac{1}{4\pi} \int \tr \Big[ bdb+2(QA+S\omega)db\Big]\qquad (21^*)
\end{equation*}
Taking into account this additional term in Eq. (21) changes the expression Eq. (24) to $\beta =0$.
It can be shown explicitly in the parton construction using the approach of Ref. \cite{Cho2014},  that correct version of Eq. (21) of Ref. \cite{Gromov2015} for the $\mathbb{Z}_k$ parafermion FQH states is explicitly given by
\begin{align}
\mathcal{L} =& \frac{M}{4\pi} \left(\frac{k}{Mk+2} A + \frac{\omega}{2} +2\alpha_{U(1)} \right) d  \left(\frac{k}{Mk+2} A + \frac{\omega}{2} + 2\alpha_{U(1)}\right) 
 + \frac{1}{4\pi} \text{tr} \left(\alpha_{SU(M)} d \alpha_{SU(M)} + \frac{2}{3} \alpha_{SU(M)}^3 \right) \nonumber\\ 
& + \frac{2k}{4\pi} \left(\frac{1}{Mk+2} A + \frac{\omega}{2} -M \alpha_{U(1)} \right) d  \left(\frac{1}{Mk+2} A + \frac{\omega}{2} -M \alpha_{U(1)} \right)  + \frac{1}{4\pi} \text{tr} \left(\alpha_{Sp(2k)} d \alpha_{Sp(2k)}  + \frac{2}{3} \alpha_{Sp(2k)} ^3 \right) \nonumber\\ 
& - \frac{c_{U(1)^{2k+M}_1}}{48\pi} \omega d \omega
\label{Eq_21}
\end{align}
Here  $\alpha_{U(1)}^\mu$, $\alpha_{SU(M)}^\mu$ and $\alpha_{Sp(2k)}^\mu$ are, respectively,  the internal $U(1)$ gauge field,   the $SU(M)$ gauge field, and   the $Sp(2k)$ gauge field, in  the  algebra of the gauge group $U(1) \times SU(M) \times Sp(2k) = U(M) \times Sp(2k)$ of the projective parton construction of the parafermion state \cite{Barkeshli2010}. The last term in Eq. (1) is the gravitational Chern-Simons term of $2k+M$ Landau levels of the parton construction.  Upon integrating out the dynamical gauge fields $\alpha_{U(1)}^\mu$, $\alpha_{SU(M)}^\mu$ and $\alpha_{Sp(2k)}^\mu$ , and carefully including the contributions of the framing anomaly to the partition functions of the gauge fields, we obtain the effective action for the spin connection $\omega_\mu$ and the external electromagnetic field $A_\mu$ to be
\begin{equation}
\mathcal{ L} = \frac{1}{4\pi} \frac{k}{Mk+2} \left(A + \frac{M+2}{2}\omega \right) d \left(A + \frac{M+2}{2}\omega \right) - \frac{c_{U(1)_{1}^{2k+M}} - c_{U(M)_{1}}- c_{Sp(2k)_1}}{48\pi}\omega d \omega
\end{equation}
which implies that  Eq. (24) of Ref. \cite{Gromov2015}
 should be replaced by  $\beta=0$.
This result is consistent with the more recent work of Bradlyn and Read \cite{Bradlyn2015} who reached this conclusion by computing the spin variance for a class of trial FQH  wave functions with the form of conformal blocks. All other results reported in Ref. \cite{Gromov2015} are correct and are not affected by this erratum.

We thank Barry Bradlyn and Nick Read for pointing out this discrepancy. This work of  was supported in part by the NSF grants No. DMR-1206790 at Stony Brook University (A.G.A.), DMR-1064319 (GYC,EF), DMR 1408713 (YY,EF) at the University of Illinois, and PHY11-25915 at KITP (EF).

%
